\documentclass[twocolumn]{jpsj3}
\usepackage{txfonts}
\usepackage{braket}

\title{Possible Odd-Frequency Superconductivity in Strong-Coupling Electron-Phonon Systems}

\def\runauthor{H. \textsc{Kusunose} {\it et al.}}
\author{Hiroaki \textsc{Kusunose}\thanks{kusu@phys.sci.ehime-u.ac.jp},
Yuki \textsc{Fuseya}$^{1}$ and
Kazumasa \textsc{Miyake}$^{1}$
}

\inst{Department of Physics, Ehime University, Matsuyama, Ehime 790-8577, Japan\\
$^{1}$Division of Materials Physics, Department of Materials Engineering Science,\\Graduate School of Engineering Science, Osaka University,\\Toyonaka, Osaka 560-8531, Japan
}

\abst{
A possibility of the odd-frequency pairing in the strong-coupling electron-phonon systems is discussed.
Using the Holstein-Hubbard model, we demonstrate that the anomalously soft Einstein mode with the frequency $\omega_{\rm E}\ll\omega_{c}$ ($\omega_{c}$ is the order of the renormalized bandwidth) mediates the $s$-wave odd-frequency triplet pairing against the ordinary even-frequency singlet pairing.
It is necessary for the emergence of the odd-frequency pairing that the pairing interaction is strongly retarded as well as the strong coupling, since the pairing interaction for the odd-frequency pairing is effective only in the diagonal scattering channel, $(\omega_{n},-\omega_{n})\to(\omega_{n'},-\omega_{n'})$ with $\omega_{n'}=\omega_{n}\gtrsim \omega_{\rm E}$.
Namely, the odd-frequency superconductivity is realized in the opposite limit of the original BCS theory.
The Ginzburg-Landau analysis in the strong-coupling region shows that the specific-heat discontinuity and the slope of the temperature dependence of the superfluid density can be quite small as compared with the BCS values, depending on the ratio of the transition temperature $T_{c}$ and $\omega_{c}$.
}

\kword{odd-frequency superconductivity, cage-like compounds, Holstein-Hubbard model}

\begin{document}
\maketitle

\section{Introduction}
A novel symmetry of superconducting pairing, which is odd in imaginary time of the relative coordinate of the Cooper pair, has been investigated for several decades.
The so-called odd-frequency (OF) superconductivity was first proposed by Berezinskii as a possible triplet pairing in the A-phase of superfluid $^{3}$He\cite{Berezinskii74}.
Two decades later, Balatsky and Abrahams revisited the singlet type of the OF pairing in the context of cuprates superconductors\cite{Balatsky92,Abrahams93,Abrahams95}.
Since then, the OF pairing has been discussed in a wide variety of theoretical models, e.g., the Kondo lattice\cite{Emery92,Coleman93,Coleman94,Coleman95,Coleman97}, the square-lattice\cite{Bulut93} and the triangular-lattice\cite{Vojta99,Yada08,Shigeta09} Hubbard, and $t$-$J$\cite{Balatsky93} models.
A universal feature of superconductivity so far stimulates to discuss the OF pairing in connection with experimental reality in the heavy-fermion systems\cite{Fuseya03,Hotta09}, superconducting junctions\cite{Tanaka07,Tanaka07a,Tanaka07b,Asano06,Asano07,Eschrig07,Tanaka08,Bergeret01,Bergeret05,Keizer06,Sosnin06,Khaire10,Sprungmann10,Anwar10}, vortex core\cite{Yokoyama08,Tanuma09}, proximity effect of superfluid $^{3}$He\cite{Higashitani09} and the cold atoms\cite{Kalas08}.

Aside from the intensive studies, the homogenous state of the OF pairing has struggled against the fundamental difficulties such as the thermodynamic instability\cite{Coleman94,Heid95,Heid95a} and the unphysical Meissner effect\cite{Abrahams95}.
These difficulties have hampered further investigation on the property of the superconducting state except a determination of the transition temperature $T_{c}$.
Recently, it is pointed out that these difficulties can be settled by the appropriate treatment of the gap function with retardation in the path-integral framework\cite{Solenov09,Kusunose10}.
As a result, the property of the OF state is ready to be examined without any theoretical ambiguity.

In the OF pairing, a strong on-site Coulomb repulsion is avoided by the off-time pairing with a vanishing equal-time pair amplitude.
In order to gain an attraction efficiently and overcome an ordinary even-frequency (EF) pairing, a strongly retarded interaction, i.e., with strong frequency dependence, as well as a repulsive part of the interaction varying slowly in frequency may be preferable.
Following this scenario, the realization of the OF pairing was examined by the mechanism of exchanging the critical antiferromagnetic fluctuation near the magnetic quantum critical point and the massless spin wave in the antiferromagnetic phase\cite{Fuseya03}.

In this paper, we focus on peculiar electron-phonon systems such as Skutterudite\cite{Sato09} and $\beta$-pyrochlore\cite{Hiroi07} families as another possible candidate for the OF pairing.
These systems are characterized by a cage-like structure with a large anharmonicity of an ionic potential where relatively isolated phonons constitute well-defined Einstein modes\cite{Tsutsui08}.
They are expected to be particularly robust against strong electron-phonon coupling inherent from the cage-like structure.
The investigations based on the Holstein-Hubbard model for such systems concluded that dynamical effect leads to the further softening of the Einstein mode and the conduction band is heavily renormalized near the half filling\cite{Meyer02,Freericks94,Freericks95}.
From these conclusions, the effective attractive interaction works in a very narrow range $\epsilon\lesssim \omega_{\rm E}$ and the strength of the pairing attraction falls into the strong-coupling region.
Namely, $\lambda_{\rm ep}\equiv\rho_{\rm F}V_{0}\sim g^{2}/\omega_{\rm E}\omega_{c}\gtrsim 1$, where $g$, $\omega_{\rm E}$ and $\omega_{c}$ are the renormalized  electron-phonon coupling, the Einstein frequency and the bandwidth, respectively, with the condition $\omega_{\rm E}\ll \omega_{c}$.
It is the situation that is quite suitable for the OF pairing, although only the possibility of the EF pairing has been so far investigated extensively\cite{Freericks94,Freericks95,Schuttler95,Pao98,Grzybowski07,Takada96,Hotta96,Hotta97,Linden95,Marsiglio90}.

The purpose of this paper is two folds: (i) to investigate the favorable conditions for the OF pairing in the strong-coupling electron-phonon systems, (ii) to elucidate the properties of the $s$-wave triplet as the simplest case of the OF pairing.
In \S2, we give the Holstein-Hubbard model and the formulation for the OF pairing state.
The discussion of the structure of the pairing interaction and the properties of the OF phase are given in \S3.
The last section is devoted to the Ginzburg-Landau description of the strong-coupling pairing, and the summary of the paper.

\section{Model and Formulation}

We consider the Holstein-Hubbard model,
\begin{align}
&H=H_{\rm ep}+H_{\rm c},
\cr
&\quad
H_{\rm ep}=\omega_{\rm E}\sum_{i}\left(b_{i}^{\dagger}b_{i}^{}+\frac{1}{2}\right)+g\sum_{i}\left(b_{i}^{}+b_{i}^{\dagger}\right)\sum_{\alpha}\left(n_{i\alpha}-\frac{1}{2}\right).
\cr
&\quad
H_{\rm c}=\sum_{\mib{k}\alpha}\xi_{\mib{k}}c_{\mib{k}\alpha}^{\dagger}c_{\mib{k}\alpha}^{}
+U\sum_{i}n_{i\uparrow}n_{i\downarrow},
\end{align}
where $n_{i\alpha}=c_{i\alpha}^{\dagger}c_{i\alpha}^{}$ is the electron density operator for the spin $\alpha$, and the Einstein phonons $b_{i}^{}$ and $b_{i}^{\dagger}$ with the frequency $\omega_{\rm E}$ couple with the electron density at the site $i$.
As we mentioned in the introduction, all the parameters are regarded as the renormalized ones.

Using the path-integral form and integrating out the Einstein phonons\cite{Bickers89}, we obtain the partition function with the effective action as
\begin{align}
&Z=\int{\cal D}\overline{c}\int{\cal D}c\,e^{-(S_{0}+S_{\rm int})},
\cr
&\quad
S_{0}=-\sum_{\alpha}\sum_{k}\left(i\omega_{n}-\xi_{\mib{k}}\right)\overline{c}_{\alpha}(k)c_{\alpha}(k),
\cr
&\quad
S_{\rm int}=\frac{T}{2}\sum_{\alpha\beta}\sum_{kk'}\sum_{q}v(k-k')\overline{c}_{\alpha}(k)\overline{c}_{\beta}(q-k)c_{\beta}(q-k')c_{\alpha}(k'),
\cr&
\end{align}
where $k=(\mib{k},i\omega_{n})$, $\sum_{k}=\sum_{\mib{k},n}$.
Here the effective interaction among electrons is given by
\begin{equation}
v(q)=U-V_{0}\frac{\omega_{\rm E}^{2}}{\epsilon_{m}^{2}+\omega_{\rm E}^{2}},
\quad
V_{0}=\frac{2g^{2}}{\omega_{\rm E}},
\end{equation}
where $q=({\mib q},\epsilon_{m})$ with the bosonic Matsubara frequency $\epsilon_{m}=2\pi T m$.

Let us examine the general filling of the electron density $n$, and no particular structure of the Fermi surface is concerned for the superconductivity.
In this case, the ordinary Cooper instability is expected to occur in the isotropic $s$-wave pairing.
Following the recent theoretical development for the OF superconductivity\cite{Solenov09,Kusunose10}, the gap equations for the $s$-wave EF singlet pairing $\Delta_{\alpha\beta}(k)=\Delta_{+}(i\omega_{n})(i\sigma^{y})_{\alpha\beta}$ and the $s$-wave OF unitary triplet pairing\cite{dvector} $\Delta_{\alpha\beta}(k)=i\Delta_{-}(i\omega_{n})\delta_{\alpha\beta}$ are given by
\begin{equation}
\Delta_{\phi}(i\omega_{n})=-T\sum_{n'=0}^{\infty}V_{\phi}(\omega_{n},\omega_{n'})F_{\phi}(i\omega_{n'}),
\quad
(\phi=\pm),
\label{gapeq}
\end{equation}
where we have used the symmetry $\Delta_{\phi}(i\omega_{n})=\phi\Delta_{\phi}(-i\omega_{n})$ of the gap functions for the EF ($\phi=+$) and the OF ($\phi=-$) pairings ($\omega_{n}=\pi T(2n+1)$ is the fermionic Matsubara frequency).
Here we have introduced the interaction matrices as
\begin{equation}
V_{\pm}(\omega_{n},\omega_{n'})=v(i\omega_{n}-i\omega_{n'})\pm v(i\omega_{n}+i\omega_{n'}).
\end{equation}
The local anomalous green function is given by
\begin{equation}
F_{\phi}(i\omega_{n})
=\frac{2\rho_{\rm F}\Delta_{\phi}(i\omega_{n})}{\sqrt{\omega_{n}^{2}+|\Delta_{\phi}(i\omega_{n})|^{2}}}
\tan^{-1}\left(\frac{\omega_{c}}{\sqrt{\omega_{n}^{2}+|\Delta_{\phi}(i\omega_{n})|^{2}}}\right).
\end{equation}
Here we have used the constant density of states, $\rho(\xi)=\rho_{\rm F}\,\theta(\omega_{c}-|\xi|)$ for the conduction electrons and the cut-off $\omega_{c}$ is the order of the renormalized bandwidth.
In order to determine the transition temperature $T_{c}$, the linearized gap equation is often used,
\begin{equation}
\lambda\Delta_{\phi}(i\omega_{n})=-2\rho_{\rm F}T\sum_{n'=0}^{\infty}\frac{V_{\phi}(\omega_{n},\omega_{n'})}{\omega_{n'}}\tan^{-1}\left(\frac{\omega_{c}}{\omega_{n'}}\right)\Delta_{\phi}(i\omega_{n'}),
\label{lingap}
\end{equation}
where the eigenvalue $\lambda(T)$ has been introduced to check the development of the Cooper instability signaled by $\lambda(T_{c})=1$.

With the help of the solution of the gap equation (\ref{gapeq}), we can express the free energy\cite{Kusunose10} (measured from that of the normal state) as
\begin{align}
{\cal F}&=-2\pi\rho_{\rm F}T\sum_{n=0}^{\infty}\frac{\left(\sqrt{\omega_{n}^{2}+|\Delta_{\phi}(i\omega_{n})|^{2}}-\omega_{n}\right)^{2}}{\sqrt{\omega_{n}^{2}+|\Delta_{\phi}(i\omega_{n})|^{2}}}
\cr
&=-2\pi\rho_{\rm F}T\sum_{n=0}^{\infty}\frac{|\Delta_{\phi}(i\omega_{n})|^{4}}{\sqrt{\omega_{n}^{2}+|\Delta_{\phi}(i\omega_{n})|^{2}}\left(\sqrt{\omega_{n}^{2}+|\Delta_{\phi}(i\omega_{n})|^{2}}+\omega_{n}\right)^{2}}.
\cr&
\label{fe}
\end{align}
The latter expression is convenient for the numerical computation due to the fast convergence of the Matsubara summation.
The entropy and the specific heat can be computed by the numerical differentiation as
\begin{equation}
S=-\frac{\partial{\cal F}}{\partial T},
\quad
C=T\frac{\partial S}{\partial T}.
\end{equation}
The superfluid density\cite{Kusunose10,AGD} is given by
\begin{equation}
n_{s}=2\pi T\sum_{n=0}^{\infty}\frac{|\Delta_{\phi}(i\omega_{n})|^{2}}{\left(\omega_{n}^{2}+|\Delta_{\phi}(i\omega_{n})|^{2}\right)^{3/2}}.
\end{equation}
Hereafter, we use the unit of energy or temperature as $1/2\rho_{\rm F}=\omega_{c}=1$.

\section{Results}

\begin{figure}[tb]
\begin{center}
\includegraphics[width=8.5cm]{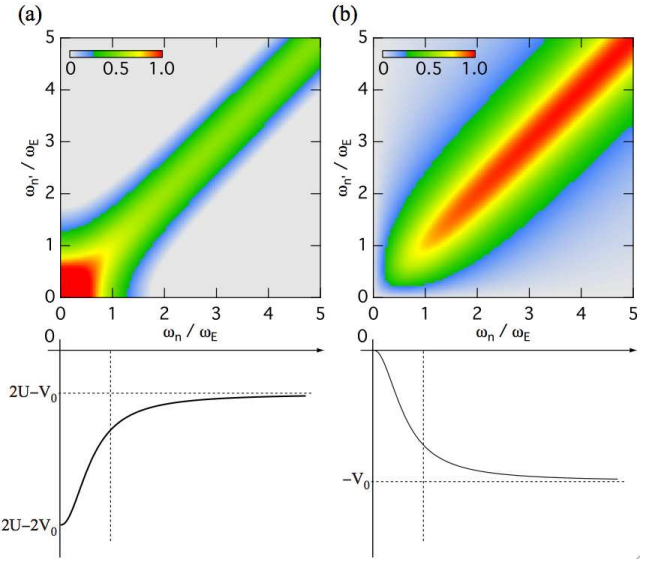}
\end{center}
\caption{(Color online) The interaction matrices for (a) the EF channel, $|V_{+}(\omega_{n},\omega_{n'})|/V_{0}$ with $U/V_{0}=0.25$, (b) the OF channel, $|V_{-}(\omega_{n},\omega_{n'})|/V_{0}$ in the upper panel. The lower panels show the diagonal parts, $V_{\pm}(\omega_{n},\omega_{n})$.}
\label{V}
\end{figure}

Let us first clarify the difference of the interaction matrices between the EF and the OF channels.
The upper panels of Fig.~\ref{V} show the intensity map of $|V_{\pm}(\omega_{n},\omega_{n'})|/V_{0}$ with $U/V_{0}=0.25$ for (a) the EF channel ($+$) and (b) the OF channel ($-$).
For the EF channel, the dominant contribution comes from the square region of $\omega_{n}$, $\omega_{n'}<\omega_{\rm E}$, and hence the summation in the gap equation (\ref{gapeq}) is usually cut off at $\omega_{n'}\sim\omega_{\rm E}$, and the diagonal part of $\omega_{n}=\omega_{n'}>\omega_{\rm E}$ is neglected.
On the other hand, the diagonal part plays an essential role in the case of the OF pairing.
Moreover, in the adiabatic limit $\omega_{\rm E}\to 0$ as consequences of strong fluctuations of ionic vibrations and/or a strong anharmonicity of the ionic potential, the diagonal part of the interaction matrices dominate the attractive interaction both for the EF and the OF pairings.

The diagonal part of the interaction matrices given by
\begin{align}
&V_{+}(\omega_{n},\omega_{n})=2U-V_{0}\left[1+\frac{1}{(2\omega_{n}/\omega_{\rm E})^{2}+1}\right],
\cr
&V_{-}(\omega_{n},\omega_{n})=-V_{0}\left[1-\frac{1}{(2\omega_{n}/\omega_{\rm E})^{2}+1}\right],
\end{align}
are shown in the lower panel of Fig.~\ref{V}.
Since the OF pairing can avoid the on-site Coulomb repulsion $U$, it is more stabilized against the EF pairing as $\omega_{\rm E}$ approaches to $0$.
At the same time, the strong-coupling condition $\lambda_{\rm ep}\equiv\rho_{\rm F}V_{0}\gtrsim 1$ is necessary for an appearance of the OF pairing because the attraction is restricted to the diagonal part of the scattering $(\omega_{n},-\omega_{n})\to(\omega_{n'},-\omega_{n'})$ with $\omega_{n'}=\omega_{n}$ and $\omega_{\rm E}\lesssim\omega_{n}\lesssim|\Delta_{-}(i\omega_{n})|\sim TV_{0}$, yielding the net amount of the attractive contributions is rather weak.

As $T$ decreases, the dominant part of the attraction ($\omega_{\rm E}\lesssim\omega_{n}\lesssim |\Delta_{-}(i\omega_{n})|$) for the OF pairing is vanishing.
Hence, the OF pairing tends to exhibit a reentrant behavior\cite{Balatsky92,Fuseya03}.
On the contrary, the dominant part of the attraction for the EF pairing becomes stronger as $T$ decreases, and it is possible to overcome the OF pairing in lower temperatures, provided the coupling constant is strong enough.

\begin{figure}[tb]
\begin{center}
\includegraphics[width=8.5cm]{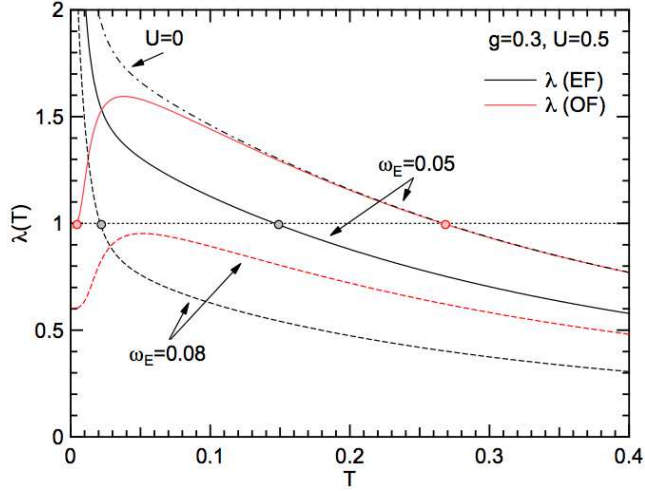}
\end{center}
\caption{(Color online) The $T$ dependence of the eigenvalue of the linearized gap equation. For $U=0$, $T_{c}$ (the open circle) for the EF pairing is slightly higher than that for the OF pairing. For $\omega_{\rm E}=0.08$, no transitions occur in the OF channel.}
\label{lambdaT}
\end{figure}

\begin{figure}[bt]
\begin{center}
\includegraphics[width=8.5cm]{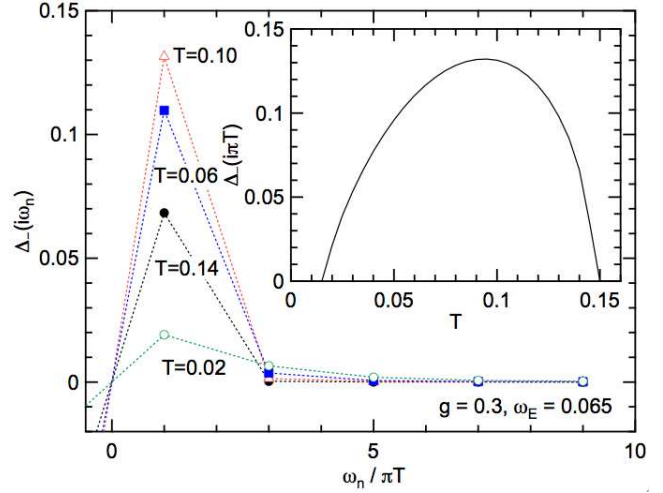}
\end{center}
\caption{(Color online) The $\omega_{n}$ dependence of the OF gap function. The inset shows the $T$ dependence of $\Delta_{-}(i\omega_{n=0})$. The dotted lines are a guide for eyes.}
\label{gap}
\end{figure}

\begin{figure}[bt]
\begin{center}
\includegraphics[width=8.5cm]{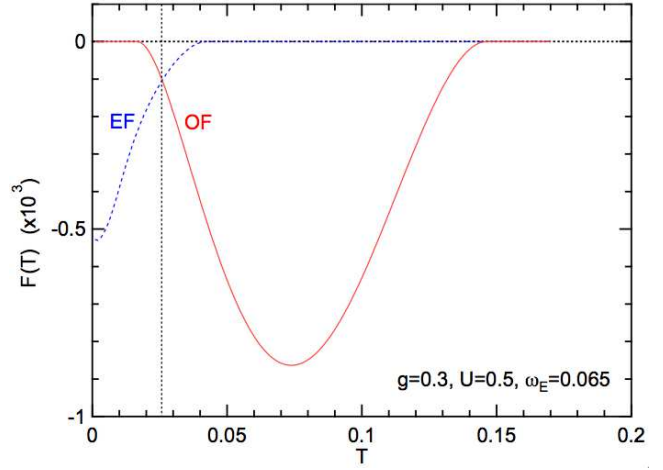}
\end{center}
\caption{(Color online) The $T$ dependence of the free energy for the OF (the solid line) and the EF (the dotted line) pairing. The vertical dotted line indicates the 1st-order transition temperature between the OF and the EF phases.}
\label{dFT}
\end{figure}

Next, we discuss the $T$-$\omega_{\rm E}$ phase diagram.
For this purpose, we solve the linearized gap equation (\ref{lingap}).
Figure~\ref{lambdaT} shows the $T$ dependence of the eigenvalues for $g=0.3$.
For $\omega_{\rm E}=0.05$ ($\lambda_{\rm ep}=1.8$) and $U=0$, $T_{c}$ for the EF pairing is slightly higher than that for the OF pairing.
When $U$ is switched on as $U=0.5$, only the EF pairing is affected to decrease $T_{c}$, and the OF pairing emerges.
As mentioned in the above, $\lambda(T)$ decreases below $T\sim \omega_{\rm E}$ due to the vanishing attraction for the OF pairing.
The reentrant lower $T_{c}$ to the normal phase appears in this case.
For $\omega_{\rm E}=0.08$ ($\lambda_{\rm ep}\simeq 1.1$), the eigenvalue does not reach unity and only the EF pairing occurs.

The solution of the gap function for the OF pairing is shown in Fig.~\ref{gap}.
Owing to the strongly retarded attraction, the gap function decreases very fast as $\omega_{n}$ increases.
The $T$ dependence of the most dominant component, $\Delta_{-}(i\pi T)$, is shown in the inset of Fig.~\ref{gap}.
It exhibits a maximum at $T\gtrsim\omega_{\rm E}$.
Note that the fast decrease of the gap function in $\omega_{n}$ is also obtained in the case of the EF pairing, and it is the characteristic feature of the strongly retarded attraction.

In order to determine a complete phase diagram, we should compare the free energy of the OF and the EF phases.
The $T$ dependence of the free energy (\ref{fe}) is shown in Fig.~\ref{dFT}.
Due to the reentrant behavior of the OF phase, there exists the 1st-order transition to the EF phase at $T\simeq 0.026$ around which we expect a coexistent phase.

\begin{figure}[tb]
\begin{center}
\includegraphics[width=8.5cm]{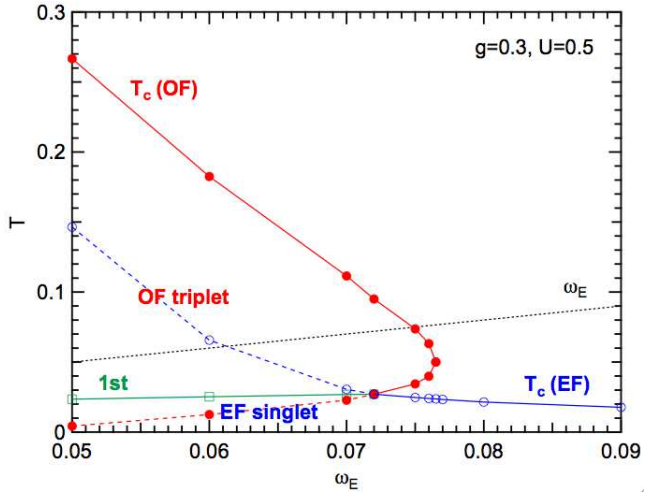}
\end{center}
\caption{(Color online) The $T$-$\omega_{\rm E}$ phase diagram. The ``1st'' in the OF triplet phase indicates the 1st-order transition to the EF singlet phase determined by the comparison of the free energies of both phases. The solid lines represent the actual transition lines, while the dotted lines indicate the fictitious transitions without the other phases.}
\label{TcOmega0}
\end{figure}

By a similar analysis, we draw the $\omega_{\rm E}$-$T$ phase diagram as shown in Fig.~\ref{TcOmega0}.
The OF phase is more stabilized when the pairing interaction becomes strongly retarded with decrease of $\omega_{\rm E}$ that simultaneously makes the coupling constant $\lambda_{\rm ep}$ to increase.
The upper $T_{c}$ is mostly higher than $\omega_{\rm E}$ which reflects the strong-coupling condition necessary for the OF pairing.
The reentrant behavior appears in a certain part of the phase diagram, although the EF pairing emerges as the low-$T$ phase in most cases.
Note that the EF phase should be suppressed as $U$ increases.
The OF phase exists only below $\omega_{\rm E}\sim 0.077$ ($\lambda_{\rm ep}\gtrsim 1.17$) indicating again that the strong coupling is necessary for the appearance of the OF pairing.

\begin{figure}[tb]
\begin{center}
\includegraphics[width=8.5cm]{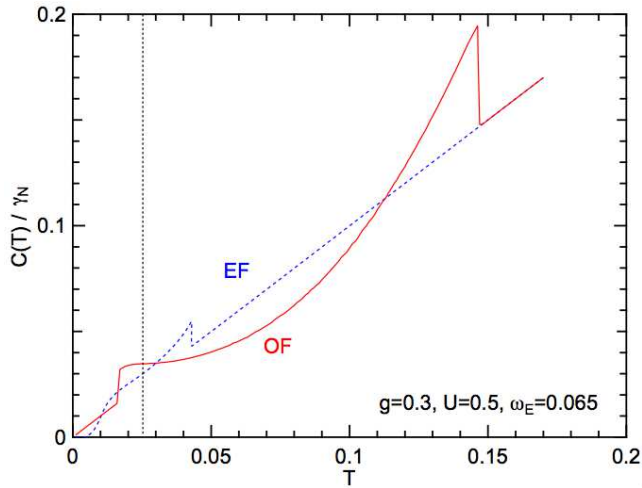}
\end{center}
\caption{(Color online) The $T$ dependence of the specific heat $C(T)/\gamma_{N}$ where $\gamma_{N}$ is the $T$-linear coefficient in the normal phase. The discontinuity at $T_{c}$ is much smaller than that of the BCS theory. The vertical line represents the 1st-order transition to the EF pairing.}
\label{CgT}
\end{figure}

\begin{figure}[tb]
\begin{center}
\includegraphics[width=8.5cm]{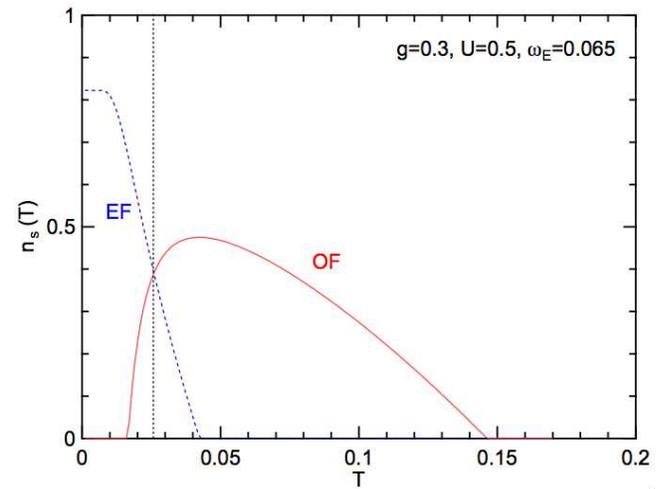}
\end{center}
\caption{(Color online) The $T$ dependence of the superfluid density, $n_{s}(T)$. Both for the EF and OF pairings, the slope of $n_{s}$ at $T_{c}$ is rather small due to the characteristics of the attractive interaction.}
\label{NsT}
\end{figure}

The $T$ dependence of the specific heat is shown in Fig.~\ref{CgT}.
The discontinuity $\Delta C(T_{c})/\gamma_{N}T_{c}\simeq0.34$ at $T_{c}\simeq0.148$ is much smaller than that of the BCS theory, $1.43$\cite{Fuseya03}.
The discontinuity for the EF pairing ($\simeq 0.24$ at $T_{c}\simeq0.042$) is also rather small.
Thus, the smallness of the discontinuity is paradoxically the consequence of the strong-coupling and the {\it strongly retarded} attraction.

Figure~\ref{NsT} shows the $T$ dependence of the superfluid density.
The $T$ dependence of $n_{s}$ near $T_{c}$, $n_{s}(T)\simeq1.18(T_{c}-T)/T_{c}$ for the OF pairing and $\simeq 0.97(T_{c}-T)/T_{c}$ for the EF pairing are also smaller than that of the BCS theory, $n_{s}(T)\simeq2(T_{c}-T)/T_{c}$.
This is also due to the strongly retarded attraction.
The relevance between the strongly retarded attraction and the smallness of the specific-heat jump and the slope of the superfluid density will be discussed below.

\section{Discussions and Conclusions}

In the previous section, we have demonstrated the necessary condition for the appearance and the characteristic features of the OF pairing.
The favorable conditions for the OF pairing are the strong-coupling {\it and the strongly retarded} attractive interaction.
These conditions result in a Cooper pair whose extent is quite small in space and large in {\it time}.
In this sense, it can be said that the OF superconductivity is the opposite limit of the one in the original BCS theory.

As was shown in the previous section, the $\omega_{n=0}$ component of the gap function plays a dominant role.
Namely, for $\omega_{\rm E}\ll T_{c}$, $\Delta_{0}\ll\omega_{c}$ and $T\lesssim T_{c}$, we adopt the Ginzburg-Landau description with the approximate gap function,
\begin{equation}
\Delta_{-}(i\omega_{n})\simeq i\Delta_{0}(\delta_{n,0}-\delta_{n,-1}),
\end{equation}
($\Delta_{0}$ is real and $\Delta_{-}(\tau)=2\Delta_{0}\sin(\pi T\tau)$ in the imaginary-time domain).
In this approximation, the $T$ dependence of $\Delta_{0}$ near $T_{c}$ from the gap equation is given by
\begin{equation}
\Delta_{0}(T)=\pi T_{c}\sqrt{\frac{T_{c}-T}{T_{c}}}\zeta^{1/2},
\quad
\zeta\equiv
\frac{4T_{c}}{\omega_{c}}=2\left(1-\frac{1}{\lambda_{0}}\right),
\end{equation}
where $\lambda_{0}=-\rho_{\rm F}V_{-}(\pi T_{c},\pi T_{c})\sim \lambda_{\rm ep}$, and $\lambda_{0}>1$.
The quantity $\zeta$ is regarded as the measure of the strong-coupling and the strongly retarded interaction.
The free energy is then obtained as
\begin{equation}
{\cal F}\simeq -2\pi\rho_{\rm F}T_{c}\frac{\Delta_{0}^{4}}{4\omega_{0}^{3}}
=-\frac{\pi^{2}}{2}\rho_{\rm F}(T_{c}-T)^{2}\zeta^{2},
\end{equation}
which gives the discontinuity of the specific heat at $T=T_{c}$ as
\begin{equation}
\frac{\Delta C}{\gamma_{N}T_{c}}=\frac{3}{2}\zeta^{2},
\quad
\left(\gamma_{N}=\frac{2\pi^{2}\rho_{\rm F}}{3}\right).
\end{equation}
The superfluid density also reads
\begin{equation}
n_{s}\simeq 2\pi T_{c}\frac{\Delta_{0}^{2}}{(\omega_{0}^{2}+\Delta_{0}^{2})^{3/2}}=2\frac{T_{c}-T}{T_{c}}\zeta.
\end{equation}
These quantities exhibit the same exponents as those of the BCS theory, but the coefficient is different by factors of certain power of $\zeta$.
It is interesting to note that the values of the BCS theory are almost reproduced by putting $\zeta\sim 1$ ($\lambda_{0}\sim 2$).
Thus, the coefficients become rather smaller when $\zeta\lesssim1$, while they become larger for $\zeta>1$ as in the case of the ordinary strong-coupling superconductivity.
We should emphasize however that these behaviors are the consequence of the strong retardation, and hence there exists little difference between the OF and the EF pairings near $T_{c}$, if the interaction is strongly retarded ($T_{c}$ for the EF pairing is obtained by replacing $V_{-}$ in $\lambda_{0}$ with $V_{+}$).
A detailed analysis for the strong-coupling (adiabatic) limit of the Eliashberg theory (for the EF pairing) is available in the literature\cite{Combescot95}.

In this paper, we have neglected both the self-energy and the vertex corrections.
Naively speaking, the self-energy correction tends to decrease the effective coupling constant $\lambda_{\rm ep}$ by a factor of the mass renormalization, and the OF phase would be suppressed.
As was shown in the previous section, the OF pairing is formed at rather high frequency $\omega_{n=0}=\pi T\gg \omega_{\rm E}$ where the mass renormalization factor could be small as compared with the low-$\omega$ limit, $1+\lambda_{\rm ep}$.
Moreover, the strong retardation and the strong coupling are the necessary conditions for the OF pairing to appear.
In this situation, it has been argued the breakdown of the Migdal theorem\cite{Aleksandrov87,Alexandrov01,Grimaldi95,Cappelluti01,Chubukov05,Anderson85,Yu84,Varma85,Kusunose96,Kusunose96a} and the vertex correction should be treated properly.
The inclusion of the self-energy and the vertex corrections are left for the future investigation.

In summary, we have discussed the possible OF superconductivity in the strong-coupling electron-phonon systems based on the Holstein-Hubbard model.
The presence of the particularly soft Einstein phonon mode provides the opposite situation of the original BCS theory, i.e., the strong-coupling and the strongly retarded attractive interaction.
They are the favorable conditions for the OF superconductivity, where the extent of the Copper pair becomes small in space and large in time.
The Ginzburg-Landau analysis for this situation concludes that the specific-heat jump and the slope of the superfluid density at $T_{c}$ are factorized by the ratio $\zeta=4T_{c}/\omega_{c}$, and they can be much smaller than those of the BCS theory for $\zeta\lesssim 1$.

\section*{Acknowledgments}
This work was supported by a Grant-in-Aid for Scientific Research on Innovative Areas ``Heavy Electrons" (No.20102008) of The Ministry of Education, Culture, Sports, Science, and Technology (MEXT), Japan.
One of the authors (K.M.) is supported in part by a Grant-in-Aid for Scientific Research on Innovative Areas ``Topological Quantum Phenomena" (No.22103003) of MEXT, and by a Grant-in-Aid for Scientific Research (No.19340099) of the Japan Society for the Promotion of Science (JSPS).
Y.F. is supported by a Grant-in-Aid for Young Scientists (No.21840035) of JSPS.


\begin{thebibliography}{99}
\bibitem{dvector} The direction of the $\mib{d}(i\omega_{n})$ vector cannot be determined in this paper. We choose $\mib{d}\parallel y$ for simplicity.
\end{thebibliography}
\end{document}